\documentclass{vgtc}                          

\ifpdf
  \pdfoutput=1\relax                   
  \usepackage{graphicx}                
  \DeclareGraphicsExtensions{.pdf,.png,.jpg,.jpeg} 
\else
  \usepackage{graphicx}                
  \DeclareGraphicsExtensions{.eps}     
\fi%

\graphicspath{{figures/}{pictures/}{images/}{./}} 

\usepackage{microtype}                 
\PassOptionsToPackage{warn}{textcomp}  
\usepackage{textcomp}                  
\usepackage{mathptmx}                  
\usepackage{times}                     
\usepackage{cite}                      
\usepackage{tabu}                      
\usepackage{booktabs}                  

\usepackage{float} 
\usepackage{subfigure}
\usepackage{tikz}
\usepackage{mathtools}
\usepackage{enumitem}
\usepackage{multicol, blindtext}
\usepackage{comment}
\usepackage{graphicx}
\usepackage[finalnew]{trackchanges}

\addeditor{JC}
\addeditor{RL}

\onlineid{0}

\vgtccategory{Research}

\vgtcinsertpkg

\title{Toward A Deep Understanding of What Makes a Scientific \\ Visualization Memorable}

\author{Rui Li\thanks{e-mail: li.8950@osu.edu} %
\and Jian Chen\thanks{e-mail: chen.8028@osu.edu}} %
\affiliation{\scriptsize The Ohio State University}

\teaser{
  \centering
  \subfigure[Selected most memorable visualizations]{
\label{teaser1}
\includegraphics[width=0.48\linewidth]{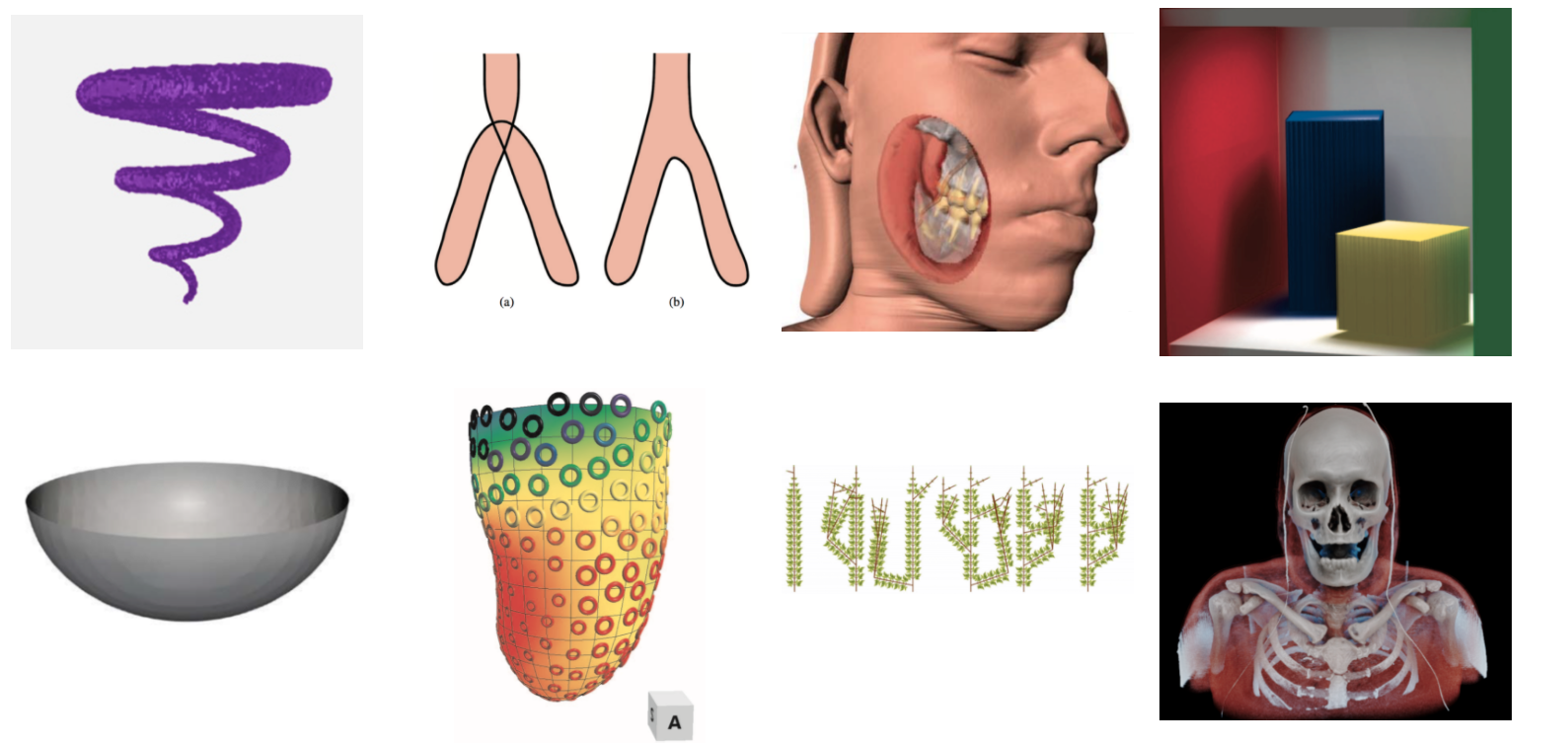}}
\hspace{1pt}\vrule\hspace{1pt}%
\subfigure[Selected least memorable visualizations]{
\label{teaser2}
\includegraphics[width=0.48\linewidth]{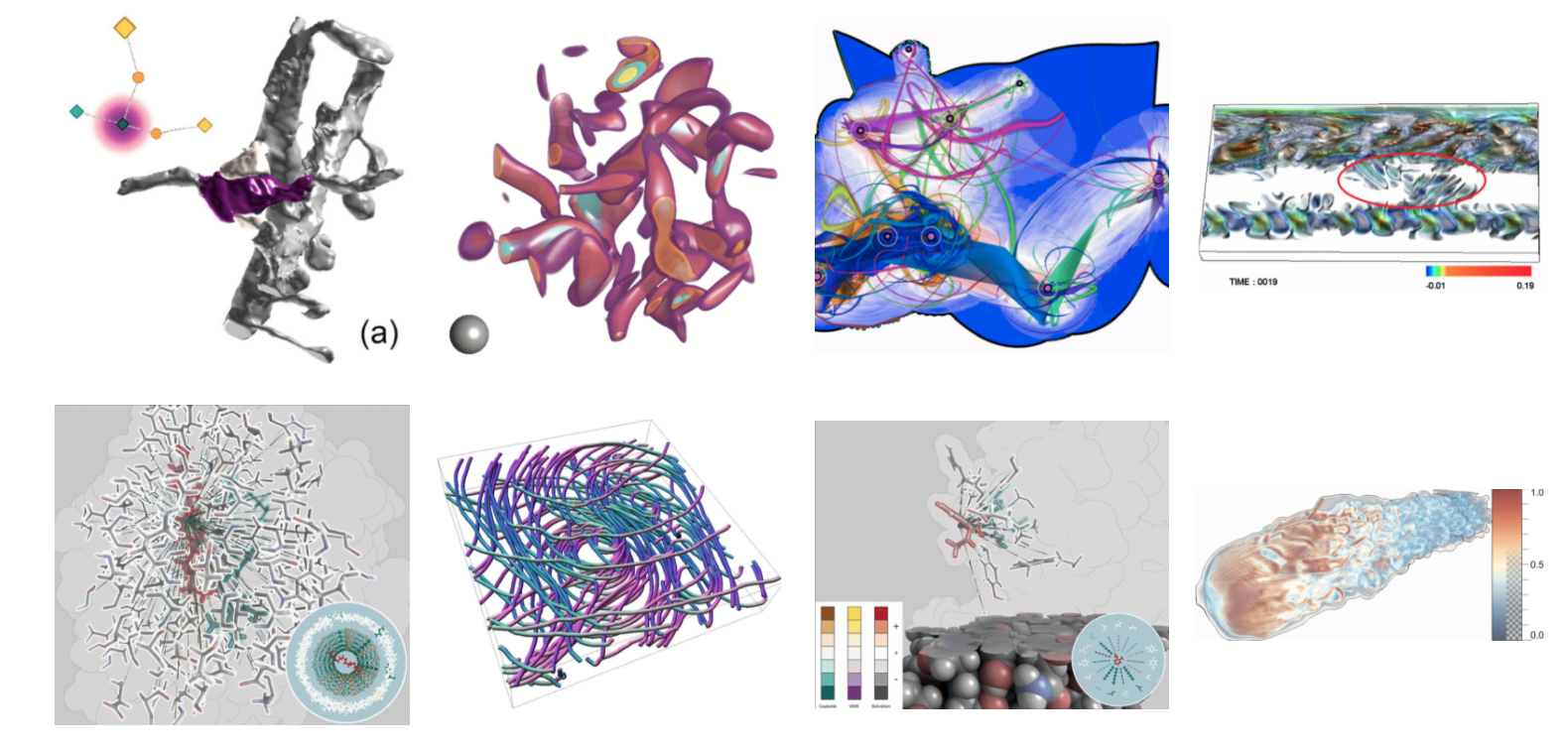}}
  \caption{Most and least memorable scientific visualizations. From (a) to (b), left to right, \add{and} top to bottom: image courtesy of Quan et al.~\cite{quan2018intelligent}, Marino and Kaufman~\cite{marino2016planar}, Rautek et al.~\cite{rautek2014vislang}, Schlegel et al.~\cite{schlegel2011extinction}, Thomas and Natarajan~\cite{thomas2014multiscale}, Meyer-Spradow et al.~\cite{meyer2008glyph}, Kolesar et al.~\cite{kolesar2017fractional}, Zhang and Ma~\cite{zhang2013lighting}, Mohammed et al.~\cite{mohammed2018abstractocyte}, Tao et al.~\cite{tao2012structure}, Sagrista et al.~\cite{sagrista2017topological}, Gu et al.~\cite{gu2016mining}, Hermosilla et al.~\cite{hermosilla2017physics}, Marchesin et al.~\cite{marchesin2010view}, \add{Hermosilla et al.{~\cite{hermosilla2017physics}}}, \add{and} Lampe et al.~\cite{lampe2009curve}.}
	\label{fig:teaser}
}

\abstract{We report results from a preliminary study exploring the memorability of spatial scientific visualizations, the goal of which is to understand the visual features that contribute to memorability.
The evaluation metrics include three objective measures (\textit{entropy}, \textit{feature congestion}, \textit{the number of edges}), four subjective ratings (\textit{clutter}, \textit{the number of distinct colors}, \textit{familiarity}, and \textit{realism}), and two sentiment ratings (\textit{interestingness} and \textit{happiness}).
We curate 1142 scientific visualization (SciVis) images from the original 2231 images in published IEEE SciVis papers from 2008 to 2017 and compute memorability scores of 228 SciVis images from data collected on Amazon Mechanical Turk (MTurk).
Results showed that the memorability of SciVis images is mostly correlated with \textit{clutter} and \textit{the number of distinct colors}. We further investigate the differences between scientific visualization and infographics as a means to \change{design more memorable scientific visualizations}{understand memorability differences by data attributes.}
} 

\CCScatlist{
  \CCScatTwelve{Memorability}{Clutter}{Color}{Affectiveness}
}

\begin{document}


\firstsection{Introduction}

\maketitle

\maketitle
The inherent complexity of scientific data has led to many innovative solutions for visualizing complex spatial phenomenon such as structural patterns and relationships. The design of these techniques is based largely on the subjective experience of visualization experts. \change{it is almost unknown}{We still have limited knowledge on} how and why visualizations are effective, what humans see and remember, and whether or not these techniques promote \change{comprehension}{engaging experiences} and learning. 

Recent studies have revealed that our brain is sensitive to a high-level property that guides memory, touched upon the concept of ~\textit{memorability},
defined
as ``\textit{an intrinsic, perceptual stimulus 
property correlated with the likelihood of an image being later remembered or forgotten}''~\cite{bainbridge2017memorability}. 
The memorability of different images shows high consistency across different groups of people~\cite{isola2014makes} and
this consistency holds for faces~\cite{bainbridge2013intrinsic}, photographs~\cite{isola2014makes}, and infographics~\cite{borkin2013makes}.
\change{However, its effect in spatial
scientific visualization has not been explored.}
{Participants have more consistent memorability experience when seeing scientific visualizations than that of information visualization{~\cite{borkin2016beyond}}.}
\add{Could memorability also be an intrinsic property of scientific visualizations? What visual features contribute the memorability of information visualization also make scientific visualization more memorable?}

The first contribution of this work is to advance our understanding of intrinsic visualization attribute of memorability. Our study first measures SciVis image memorability and then reveals whether memorability is correlated with 3D visualization image features in the objective and subjective metrics\remove{(e.g., happiness and interestingness)}. 
We adapt and expand recent vision science studies~\cite{dasgupta2017familiarity,saket2016comparing} and visualization methods and measured three objective \add{(entropy, feature congestion, and the number of edges),} \change{and six subjective}{four subjective (clutter, number of distinctive colors, familiarity, and realism), and two sentiment (interestingness or happiness)} metrics to investigate the relationship between memorability and \change{other subjective properties}{these image features}.

The second contribution of this work is a\remove{n} scientific visualization memorability database.
This database \change{should}{could} ultimately promote the understanding of the evolution of spatial data analysis and be useful for benchmarking visualization methods. The current image collection contains 2231 original SciVis images and 1142 curated SciVis memorability dataset of which 228 are annotated with their memorability scores, all accessible at https://ivclexp.github.io/scivismemorability/.

\section{Related work}
This section discusses studies closely related to our work in memorability and \change{image}{visual} features. 

\subsection{Memorability of Visualizations And Feature Space}

Visualization aims to present data to aid communication and help transform insights into knowledge. 
\add{Understanding the transition steps from human visual perception and memory is crucially important because we have to know what we see before we
understand, remember, and interact with data~\cite{bainbridge2017memorability}.}
\remove{
} 

Many features from low-level to high-level are found correlated with memorability \add{in images and information visualization}.
Low-level features are used to describe image elements such as color statistics and luminance.
For example, Borkin et al.'s online crowdsourcing experiment reported that \remove{visual attributes like}
color, density, and data-ink ratio enhance infographics memorability~\cite{borkin2013makes}.
However, \change{these}{some} features cannot be applied in SciVis, because data \add{from} infographics are mostly discrete, in contrast to continuous scientific \change{visualizations}{data} related to variety of physical phenomenons\remove{{~\cite{card1999readings}}}. For example, the concept of data-ink ratio suitable to measure density in information visualization would be ill-defined in spatial data in terms of legibility~\cite{chen2012effects}.
Though pictograms outperformed plain charts and texts in attracting viewers' attention{~\cite{haroz2015isotype}} and led to higher memorability scores {~\cite{borkin2016beyond}}, text displays also have many forms in spatial data visualizations~\cite{chen2004testbed}~\cite{tamura1978textural}.

In this work, we have chosen SciVis-relevant measurements. For example, 
\textit{entropy} can describe the ``busyness'' of a visualization and measures how difficult a visualization would be to compress~\cite{wang2011information}.
\change{Edge}{Edges} \remove{detection aims to find}{can define} the boundaries of objects within \change{an image}{a visualization, and} the number of edges can help estimate how many different areas a \change{picture}{visualization} contains. 
We have also measured \textit{visual clutter} defined as the extent to which there is no room in feature space for adding new salient items, followed the \textit{feature congestion} model by Rosenholtz et al.~\cite{rosenholtz2005feature}.





\subsection{High-Level and Affectiveness Features}


Memorability is often studied by presenting a brief glimpse of images in a fraction of a second. This brief exposure to an image or art work is long enough to guide human attention to important regions, reflecting viewers' personality traits or 
the inherent image attributes. Memory recall instead is studied in prolonged uses of interactive systems.
High-level features containing Gestalt groupings can help humans interpret\remove{specific} objects in complex spatial data visualization~\change{{~\cite{zhang2003low}}}{{~\cite{zhao2017bivariate}}}. 
Symmetry, alignment, collinearity, and axis orthogonality align with human cognitive processes Network diagrams are early for network layout~\cite{marriott2012memorability}. 

Affectiveness, which can be measured using sentiment analysis, can affect human creativity, trust, and analytical capacities. The affectiveness or emotional features of visualizations are often studied through interactive exploration. For example, 
Dasgupta et al.{~\cite{dasgupta2017familiarity}} investigated the relationship between familiarity of the analysis medium and domain experts' trust. Their results indicated that the visual analytic system can inspire greater trust than other media for complex tasks. Saket et al.{~\cite{saket2016comparing}} compared the enjoyment of node-link and node-link-group visualization and found the latter to be more enjoyable. Previous studies has found that memorability is distinct from other stimulus properties and is uncorrelated with aesthetics and affectiveness~\cite{isola2014makes}.

Journalists and artists creating infographics are undoubtedly capable of manipulating multiple perspectives such as colors and lighting deliberately to 
create memorable experiences. Visualization can be beautiful or artistic
to facilitates communication and interaction between the viewer and visualization and 
between the craft of design and the final product. 
But an important difference between images in vision science and visualizations is that 
visualization design is a process that proceeds from framing a problem (analyses) to a solution process (synthesis).
It is indeed hoped that viewers of visualizations would discount their emotion and generate consistent knowledge from data.

\section{Experiment}

This section describes our hypotheses, the scientific visualization dataset collection, measurement metrics, and our online crowdsourcing experiment on MTurk.

\begin{figure}[t!]
\centering  
\subfigure[]{
\label{fig:removedMulti}
\includegraphics[width=0.40\linewidth]{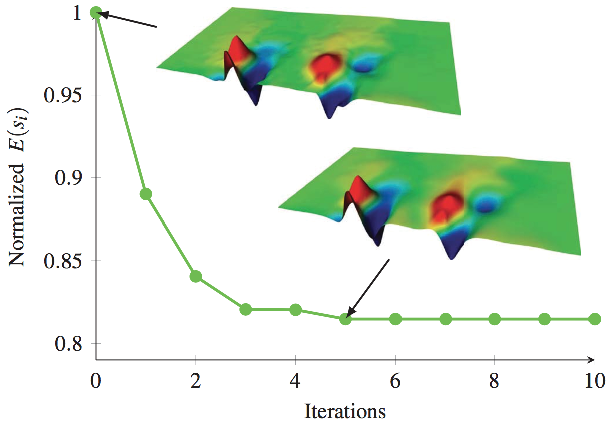}}
\subfigure[]{
\label{fig:splitMulti}
\includegraphics[width=0.55\linewidth]{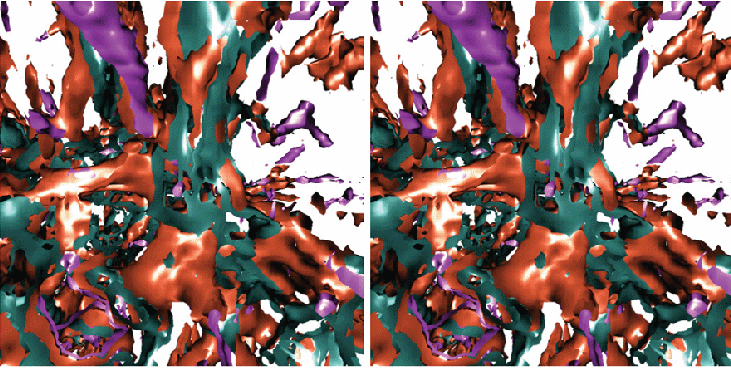}}
\caption{Curation of Multiple Panels: (a) graphs with two superimposed scientific visualizations are removed; (b) multiple-view visualizations are split. Images courtesy of G{\"u}nther et al.~\cite{gunther2014fast}, Wu et al.~\cite{wu2017direct}.}
\label{fig:multi}
\end{figure}

\subsection{Hypotheses}

Inspired by this literature and our own experiences in the design and evaluation of visualization, our hypotheses entering
the experiment were:

\begin{itemize}[noitemsep,nolistsep]
\item{\textbf{H1.}} Memorability may be intrinsic to scientific visualizations.

\item{\textbf{H2.}} A visualization may be easier to remember if it is less cluttered. 
Simple visualization (i.e., clear axes, clearly oriented) may be easier to memorize.

\item{\textbf{H3.}} Emotional responses may not be correlated with scientific visualization memorability.
\end{itemize}

\subsection{Scientific Visualization Data Collection}

\begin{figure*}[ht]
 \centering 
 \includegraphics[width=\textwidth]{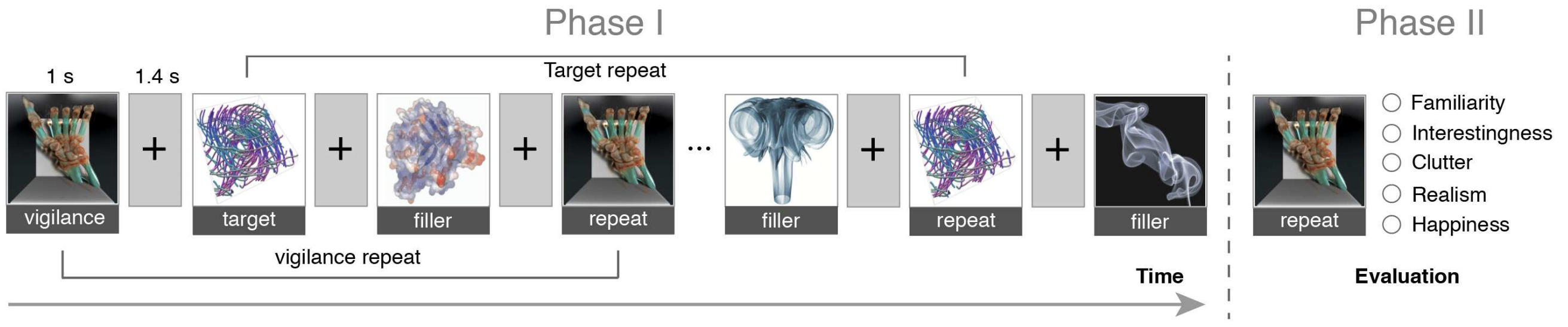}
 \caption{The online scientific visualization memorability game. Each image is shown for one second, separated by a blank interval of 1.4 seconds. After this memory game, participants are asked to fill out the evaluation form. Images courtesy of Magnus and Bruckner~\cite{magnus2018interactive}, Marchesin et al.~\cite{marchesin2010view}, Krone et al.~\cite{krone2009interactive}, Hummel et al.~\cite{hummel2010iris}, and Funck et al.~\cite{von2008smoke}}.
\label{fig:expMethod}
\end{figure*}

There are 2231 images 
appearing at the annual IEEE \change{Scivis}{SciVis} \change{conference}{papers} 
from 2008 to 2017. To ensure that the dataset contains only typical scientific visualizations (visualizations used to present continuous and spatial data), we curated \change{scientific visualization memorability}{this} data collection according to the following rules:

\begin{itemize}[noitemsep,nolistsep]
  \item Remove charts (e.g., bar and line charts) mainly appearing in evaluation papers to show statistical results.
For example, Fig.~\ref{fig:removedMulti} is removed because the two superimposed visualizations show algorithmic validation results.
  \item Remove tables, algorithms, or workflow diagrams mainly appearing in application and algorithm papers. 
  \item Split up multiple visualizations - those in which multiple images are grouped together - unless the image is part of a tool interface (Fig.~\ref{fig:multi}). For example,  Fig.~\ref{fig:splitMulti} is split up because it contains multiple single visualizations. We randomly chose one visualization and here the visualization on the right in our dataset.
\end{itemize}

Following this process, we have curated 1142 samples. 


\subsection{Metrics}

To study \change{their}{feature} correlations with scientific visualization memorability, we include a set of measures that have been shown to be useful either in \remove{similar} image and \add{information visualization} evaluation tasks \remove{in order} or in interactive explorations to support recall.


We compute three objective measures of \textit{number of edges}, \textit{feature congestion}~\cite{rosenholtz2007measuring}, and \textit{entropy}~\cite{wang2011information}.
We also collect subjective SciVis image ratings of 
(1) \textit{clutter} (Is this visualization of a neat space or a cluttered space?);
(2) \textit{familiarity} (Do the objects in the visualization look familiar to you?~\cite{saket2016comparing});
(3) \textit{realism} (Is the style of this visualization realistic or abstract? {~\cite{viola2017pondering}});
(4) \textit{happiness} (How happy does this visualization make you?~\cite{saket2016comparing}); and
(5) \textit{interestingness~\cite{dhar2011high}} (Is the visualization interesting?)
These five measures are collected using MTurk.
Three Ph.D. students studying visualization annotated \textit{the number of distinct colors} in each SciVis image on a scale of 1 to 7.

\subsection{Material and Methods}

\change{The}{Our} method\remove{that we use} to measure the memorability of scientific visualizations closely follows that of \change{Isola}{Borkin} et al.~\cite{borkin2013makes} in an online memory game on MTurk using our SciVis data collection.
\footnote{Our empirical study is hosted at: http://18.217.0.175:3000/.}
Our memorability game experiment had two phases (Fig.~\ref{fig:expMethod}). 
In the first phase, a stream of images was presented on the screen, each shown for 1 second(s), with a 1.4s-gap between consecutive images. The task of 
our participants was to press the space-bar immediately when they saw an repeated image. They received feedback on their answer \add{right} after they pressed the space-bar.

The participants could complete up to five levels with 600 different images.  Each level of the memory game had a total of 120 images and took about about 5 minutes(min) to complete. Among these 120 images, $30\%$ are targets and $70\%$ are fillers.  The targets were the images whose memorability were measured. Each target image was shown after at least 91 images and at most 109 images. 
Some fillers merely fill the space between the target images. Others are used for vigilance repeats intended to screen out participants who were not paying enough attention to the task. The vigilance repeats repeat at spacings of seven images or less and are meant to be easy to detect. Participants whose failed more than $50\%$ in the last 10 vigilance tests were blocked from the study. At the end of this phase I, participants were shown their hit rate and then went on to the next phase.

The second phase of our game is to collect subjective measures from our participants. For each target image that a participant saw during the game, we asked participants to rate five visualization attributes (clutter, familiarity, realism, happiness, and interestingness) on a 7-point Likert-scale using a range slider.


\remove{We asked participants to rate these five subjective metrics on a 7-point Likert-scale using a range slider.}

\textbf{Validation Experiment.} Before carrying out our experiment, we first performed a validation experiment 
based on Borkin et al.~\cite{borkin2013makes} to ensure that our setting is valid. We randomly choose 50 target images from their dataset and collected on average 45 responses for each image, replicating Borkin et al's experiment~\cite{borkin2013makes}. We obtained a rank correlation of 0.69 when comparing the memorability scores of these 50 images with theirs. This result confirmed that our experiment setting \change{is}{was} \change{valid}{comparable} for collecting visualization memorability scores.

\section{Results and Discussion}

This section presents the study results for 677 MTurk participants, our three \change{subjective}{objective}\remove{measures}, four subjective, and two sentiment measures, and their Spearman's rank-order correlations to memorability scores. We \change{further}{have} compared our results to \change{Borkin's}{those of Borkin et al.{~\cite{borkin2013makes}}} to leverage the differences in memorability \change{scores generated in these two sister domains}{characteristics between infographics and sciVis images}.


\subsection{MTurk Objective Measure: Memorability Scores}

\add{We collected 62 responses for each target on average in phase I and 34 responses on average in Phase II. All data \change{are}{were} used in the analysis.} 
Fig.{~\ref{fig:teaser}} shows the top eight most memorable and least memorable visualizations and Table~\ref{table:featureScores} shows the Spearman's rank-order correlation results of correlating measures to the memorability scores in this and Borkin et al.'s studies~\cite{borkin2013makes}.

We follow Borkin et al.~\cite{borkin2013makes} in using hit rate ($HR$) and false-alarm rate ($FAR$) to compute the memorability score of an image. 
The hit rate is the rate at which users give correct feedback when they see repeated images. The false-alarm rate ($FAR$) indicates the proportion of users who mistakenly identify an image as a duplicate when it appears for the first time. Further, we take both $HR$ and $FAR$ into consideration and compute $d^\prime$ \add{as an image's memorability score}, where $d^\prime=Z(HR) - Z(FAR)$ and Z is the inverse cumulative Gaussian distribution\remove{as an image's memorability score}. A high memorability score \remove{will} requires the image's $HR$ to be high and the $FAR$ to be low.

Our first hypothesis, memorability is an intrinsic in scientific visualizations, was supported.
We obtained a mean $HR$ of $48.4\%$ ($SD$ = $15\%$) and mean $FAR$ of $7\%$ ($SD$ = $5\%$).
We calculated the human consistency of our experiment and obtained a correlation rank of \add{$0.74$ for HR, $0.65$ for FAR, and $0.70$ for $d^\prime$}, averaged over 25 random half-splits.  Both our $HR$  and $FAR$ \change{was}{were} slightly lower than 
Borkin et al.'s memorability study results \add{{~\cite{borkin2013makes}}} on infographics (mean($HR$)=\change{55.36}{0.55} and mean($FAR$) = \change{13.17}{0.13}). This result demonstrated that sciVis images in our dataset were harder to remember but less prone to false alarm.

\begin{table}[!tp]
\centering
\caption{Spearman's rank-order correlations \add{($\rho$)} \remove{with ground-truth} of visualization features and their \remove{visualization} memorability scores}
\label{table:featureScores}
\resizebox{\columnwidth}{!}{%
\begin{tabular}{@{}ccccc@{}}
\toprule
\change{Feature}{Index} & Visualization feature          & $\rho$   & $\rho$ (Borkin et al.~\cite{borkin2013makes}) \\ \midrule
             Objective measures & & & \\
            \hline
f1 & \change{edges number}{number of edges}                 & -0.18 & 0.24                   \\
f2      & feature congestion           &  -0.16 & 0.05                   \\
f3      & entropy                      &  -0.14 & 0.53                   \\

\hline
    Subjective measures &&& \\
            \hline
f4      & neat space vs. clutter space &  -0.33 & /                      \\
f5      & number of distinct colors    &  -0.26 & 0.32       \\            
f6      & familiarity                  &  0.17  & /                      \\
f7      & abstract vs. realism         &  0.14 & /                      \\
\hline
    Sentiment measures &&& \\
            \hline
f8      & Interestingness              &  -0.10 & /                      \\
f9  & Happiness                    &  -0.05 & /                      \\
\bottomrule
\end{tabular}%
}
\end{table}

We are cautious about generalizations on memorability without further exploration. 
The data attributes might have contributed this difference between
scientific visualization and infographics.
The dataset used in Borkin et al's work~\add{{~\cite{borkin2013makes}}} contained a large number of infographics collected from government documents and newspapers with fairly consistent layouts and color schemes.
In contrast, 
\add{all images in our dataset are}
sciVis images, which seem to be more diverse in 
visual complexity, users' familiarity, and color schemes (many are specific to particular scientific disciplines, e.g., medical imaging or flow field), and thus \change{humans}{MTurk participants} \change{can}{may} memorize these images, leading to our low false-alarm rate. With this diversity, it might \add{also} be hard for workers to remember them from a brief exposure without considerable visualization literacy.
 

\subsection{Measurement of Visualization Features}

Overall, the low correlations between memmorability scores and SciVis image features may suggest that memorability may be an independent trait.

\textbf{Clutter, Edges, and Entropy.}
We measured both subjective and objective clutter, number of edges, and entropy.
We found a negative correlation between \textit{feature congestion} (objective clutter) and memorability, meaning that those images with high feature-congestion scores tend to be less memorable. 
However, this correlation was not statistically significant. 
As it shown in Fig. 5(a) and Table~\ref{table:featureScores}, we also found that the subjective evaluation of clutter has a correlation of 0.33 with memorability. 


The entropy was computed by the Shannon entropy equation: $Entropy = -\sum_{i=1}^{n}P_ilog_2P_i$, where $P_i$ is the probability that the difference between two adjacent pixels is equal to $i$ and $n$ is the total scale in gray-color space. \remove{In our study,}\change{w}{W}e got a correlation rank value of $-0.14$ ($p<0.001$), again a negative correlation, i.e., a visualization that contains more information is harder to remember.

These results may agree with our second hypothesis: improved spatial layout may not improve memorability.  On the one hand, the lack of significance of the objective clutter may be because our feature-congestion algorithms compute line orientation and contrast, which has limited ability to describe the complexity of the visualization. On the other hand, we may not have found a meaningful layout measurement yet to generate meaningful results.
\remove{some visualizations may contain a large number of items or lines, but have a regular distribution and thus won't be considered cluttered.}





\begin{figure}[!tp]
 \centering 
\includegraphics[width=\columnwidth]{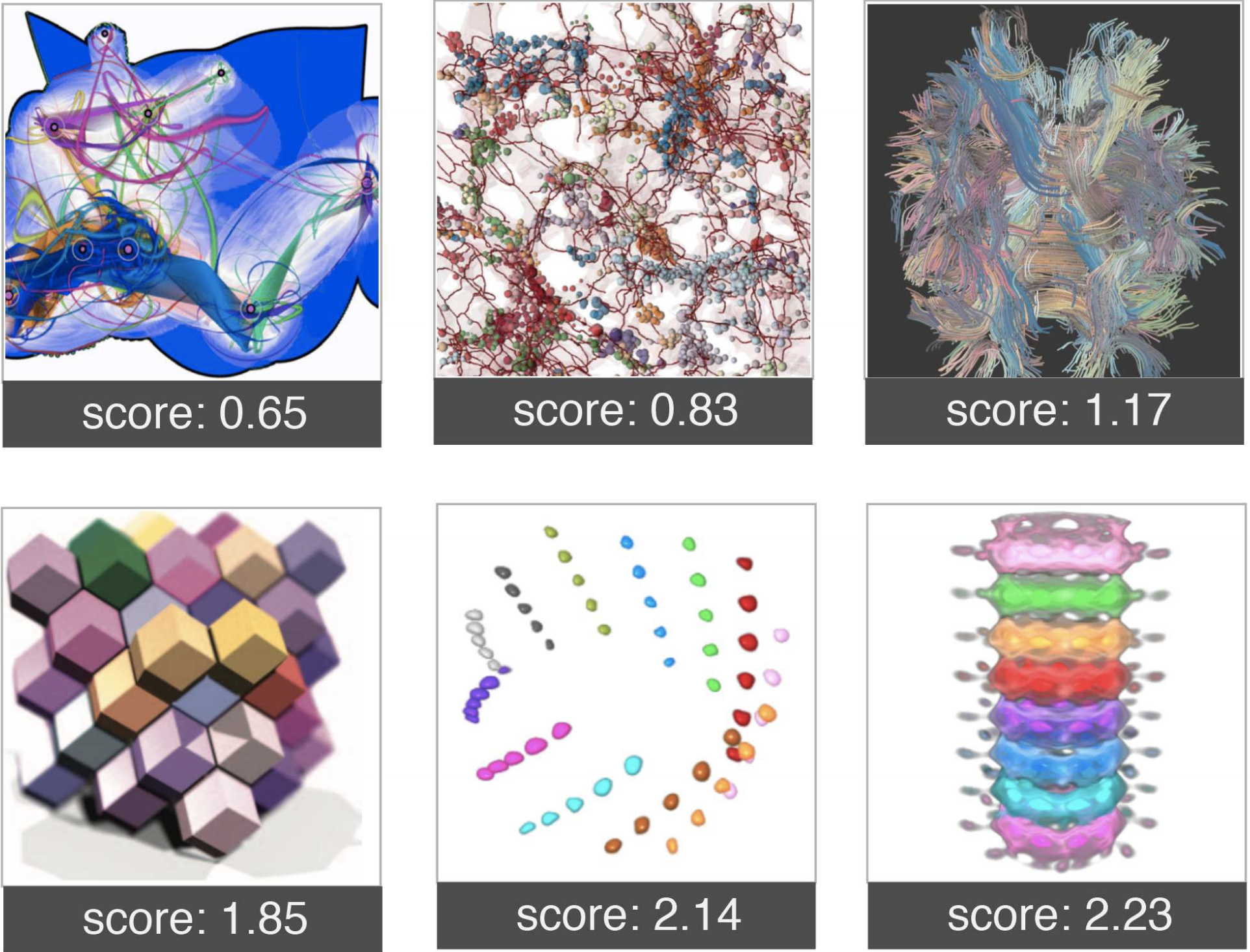}
 \caption{Colorful visualizations and clutter vs. memorability scores.
Images courtesy of Sagrista et al.~\cite{sagrista2017topological}, Mohammed et al.~\cite{mohammed2018abstractocyte}, Demiralp et al.~\cite{demiralp2009coloring}, Mirzargar and Entezari\cite{mirzargar2011quasi}, Thomas and Natarajan~\cite{thomas2013detecting}, and \add{Thomas and Natarajan}~\cite{thomas2013detecting}.}
\label{fig:memScores}
\end{figure}

\textbf{Number of Distinct Colors.} 
\remove{The first subjective measurement metric is \change{colorfulness}{the number of distinct colors}:}
Visualizations with more distinct colors (subjective judgment) had {\change{higher}{lower}}
scores than images of a single hue.  
The {Spearman's} rank-order correlation between \textit{the number of distinct colors} and memorability score was $-0.26$  and this negative correlation was significant ($p<0.001$). This result means that \remove{more colorful} images \add{with more hues} are harder to remember.

This result was at first considered contradictory to the information visualization study~\cite{borkin2013makes} indicating that colorful visualizations were easier to remember. One possible explanation of many-color-low-memorability might be the `clutter' effect: the organization of the colors leads to a degradation of performance in remembering images~\cite{rosenholtz2005feature}. This effect can be observed in Figures~\ref{fig:teaser} and ~\ref{fig:memScores}: visualizations containing more distinct colors in a less structured fashion tend to be scored lower on memorability than those colorful and well-structured images with higher memorability scores. \remove{This contrasts with the infographics used in Borkin et al., in which more colorful images tend to be more structured.}

\textbf{Familiarity and Realism.} Familiarity and realism had a weak correlation with the memorability scores for scientific visualizations (Table~\ref{table:featureScores}). 
One interesting yet not surprising finding when comparing the experiment result of our dataset and Borkin et al's~\cite{borkin2013makes} was that visualizations containing faces had high memorability scores. 
Our results again showed that `face is special.'  In addition, as shown in Fig.5(f), \textit{realism} was strongly correlated with familiarity ($\rho = 0.75, p<0.001$). 
This indicated that users tend to view visualizations that contain familiar objects as more realistic.

\textbf{Sentiment scores.}
Our third and last hypothesis was also supported. Not surprisingly,
happiness and interestingness are not statistically correlated with the memorability of scientific visualizations. 

\begin{figure}[!tp]
 \centering 
 \includegraphics[width=\columnwidth]{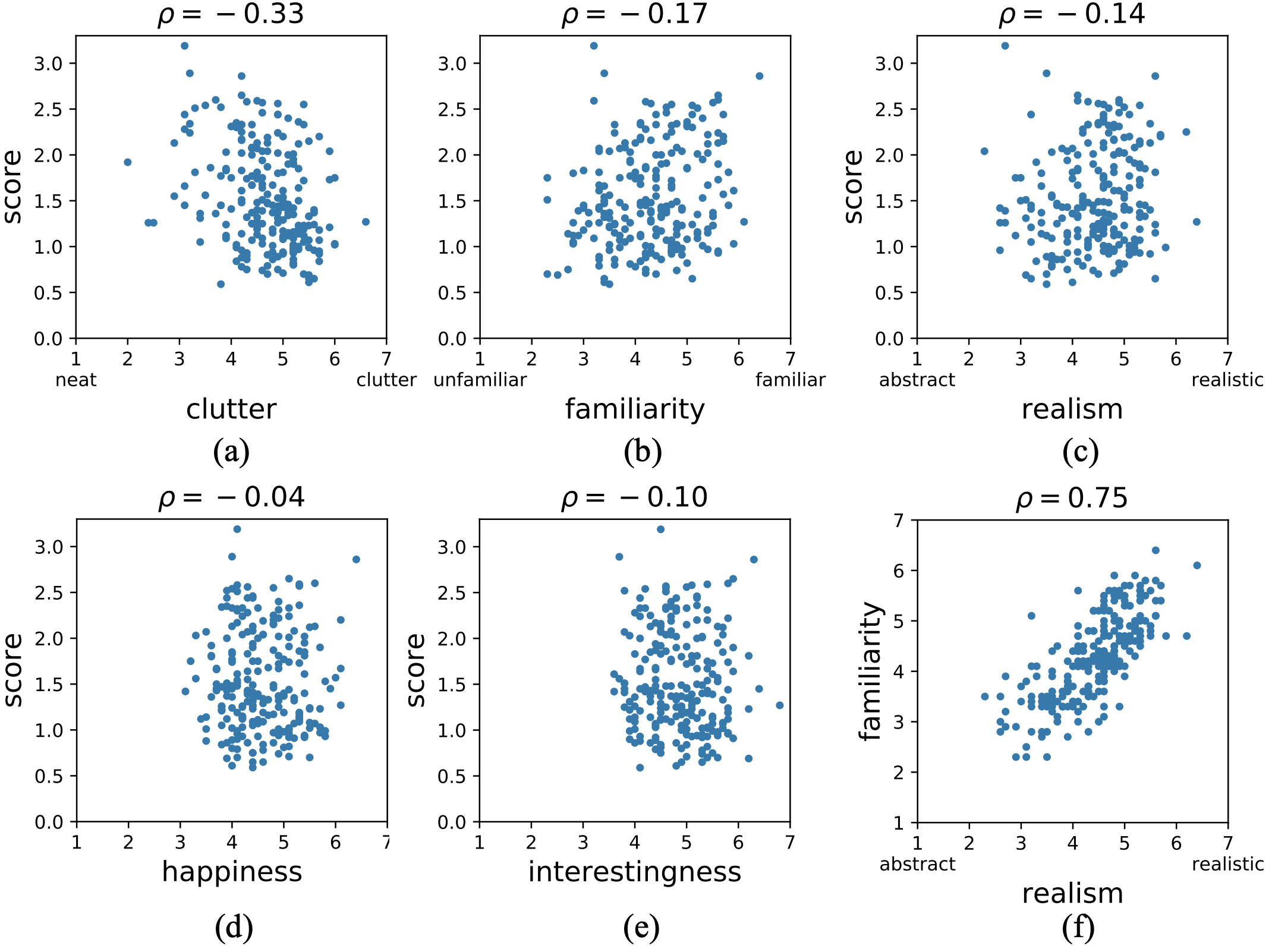}
 \caption{\remove{Subjective affective features vs. memorability score.}(a)-(c) Subjective metrics vs. memorability scores. (d)-(e) Sentiment metrics vs. memorability scores. (f) realism vs. familiarity.  Each dot represents one visualization.}
\end{figure}

\subsection{Domain Specificity vs. Generalization}

\add{Though memorability is likely to be an intrinsic attribute of scientific Visualizations, as shown in our study, 
we are cautious about generalizations on memorability without further exploration. This is because scientific data visualizations tend to be domain specific, so that the memorability of medical imaging visualizations and atmospheric visualizations may depend on domain expertise. Our current study is also limited by the type of scientific visualizations we could find. These specific visualizations often represent work in a research domain that requires years of advanced study. As a result, there could also be variance in memorability scores due to research domains, data types, and viewer' experiences. The memorability of visualizations from different domains also varies with different discovery tasks. In this case, it’s essential to ask the question, memorable to what group of experts? or in terms of what specific dataset? In future study we hope to gain further understanding of memorability and individual differences and investigate which features are important to specific data types.
}

\section{Conclusion}

We have shown that the memorability is an intrinsic property of scientific visualizations.   
The computational result of our experiment shown that visualizations with multiple colors and cluttered layouts tend to be harder to remember, contradicting previous studies in infographics. 
This result is worth further investigation to truly understand the differences between scientific and information visualizations. An interesting observation from this work is that scientific visualizations present continuous coloring schemes to encoding information.
Thus, for visual images containing the same number of distinct colors, scientific visual images may contain more \textit{unstructured} colors, which could lower memorability. Finally, our study does not contain interaction: the projected 2D images retain the visualization features shown in the original paper but not necessarily the communicative power of the original design.

\acknowledgments{
The authors would like to thank Drs. Jeremy Wolfe and Aude Oliva for discussions on memorability and the anonymous reviewers for their constructive comments. This work was supported in part by NSF IIS-1302755, CNS-1531491, and DBI-1260795 and by NIST MSE-70NANB13H181. 
Any opinions, findings, and conclusions or recommendations expressed in this material are those of the authors and do not necessarily reflect the views of National Institute of Standards and Technology (NIST) or the National Science Foundation (NSF.)
}

\bibliographystyle{abbrv-doi}

\bibliography{scimemo}

\begin{thebibliography}{10}

\bibitem{bainbridge2017memorability}
W.~A. Bainbridge, D.~D. Dilks, and A.~Oliva.
\newblock Memorability: A stimulus-driven perceptual neural signature
  distinctive from memory.
\newblock {\em NeuroImage}, 149:141--152, 2017. doi: {{%
10\hspace{.1pt}\discretionary{.}{%
}{.}\hspace{.4pt}1016\discretionary{/}{%
}{/}j\hspace{.1pt}\discretionary{.}{%
}{.}\hspace{.4pt}neuroimage\hspace{.1pt}\discretionary{.}{%
}{.}\hspace{.4pt}2017\hspace{.1pt}\discretionary{.}{%
}{.}\hspace{.4pt}01\hspace{.1pt}\discretionary{.}{%
}{.}\hspace{.4pt}063}}


\bibitem{bainbridge2013intrinsic}
W.~A. Bainbridge, P.~Isola, and A.~Oliva.
\newblock The intrinsic memorability of face photographs.
\newblock {\em Journal of Experimental Psychology: General}, 142(4):1323, 2013.
  doi: {{%
10\hspace{.1pt}\discretionary{.}{%
}{.}\hspace{.4pt}1037\discretionary{/}{%
}{/}a0033872}}


\bibitem{borkin2016beyond}
M.~A. Borkin, Z.~Bylinskii, N.~W. Kim, C.~M. Bainbridge, C.~S. Yeh, D.~Borkin,
  H.~Pfister, and A.~Oliva.
\newblock Beyond memorability: Visualization recognition and recall.
\newblock {\em IEEE TVCG}, 22(1):519--528, 2016. doi: {{%
10\hspace{.1pt}\discretionary{.}{%
}{.}\hspace{.4pt}1109\discretionary{/}{%
}{/}TVCG\hspace{.1pt}\discretionary{.}{%
}{.}\hspace{.4pt}2015\hspace{.1pt}\discretionary{.}{%
}{.}\hspace{.4pt}2467732}}


\bibitem{borkin2013makes}
M.~A. Borkin, A.~A. Vo, Z.~Bylinskii, P.~Isola, S.~Sunkavalli, A.~Oliva, and
  H.~Pfister.
\newblock What makes a visualization memorable?
\newblock {\em IEEE TVCG}, 19(12):2306--2315, 2013. doi: {{%
10\hspace{.1pt}\discretionary{.}{%
}{.}\hspace{.4pt}1109\discretionary{/}{%
}{/}TVCG\hspace{.1pt}\discretionary{.}{%
}{.}\hspace{.4pt}2013\hspace{.1pt}\discretionary{.}{%
}{.}\hspace{.4pt}234}}


\bibitem{chen2012effects}
J.~Chen, H.~Cai, A.~P. Auchus, and D.~H. Laidlaw.
\newblock Effects of stereo and screen size on the legibility of
  three-dimensional streamtube visualization.
\newblock {\em IEEE TVCG}, 18(12):2130--2139, 2012. doi: {{%
10\hspace{.1pt}\discretionary{.}{%
}{.}\hspace{.4pt}1109\discretionary{/}{%
}{/}TVCG\hspace{.1pt}\discretionary{.}{%
}{.}\hspace{.4pt}2012\hspace{.1pt}\discretionary{.}{%
}{.}\hspace{.4pt}216}}


\bibitem{chen2004testbed}
J.~Chen, P.~S. Pyla, and D.~A. Bowman.
\newblock Testbed evaluation of navigation and text display techniques in an
  information-rich virtual environment.
\newblock In {\em Proceedings of IEEE Virtual Reality}, pp. 181--188, 2004.
  doi: {{%
10\hspace{.1pt}\discretionary{.}{%
}{.}\hspace{.4pt}1109\discretionary{/}{%
}{/}VR\hspace{.1pt}\discretionary{.}{%
}{.}\hspace{.4pt}2004\hspace{.1pt}\discretionary{.}{%
}{.}\hspace{.4pt}1310072}}


\bibitem{dasgupta2017familiarity}
A.~Dasgupta, J.-Y. Lee, R.~Wilson, R.~A. Lafrance, N.~Cramer, K.~Cook, and
  S.~Payne.
\newblock Familiarity vs trust: A comparative study of domain scientists' trust
  in visual analytics and conventional analysis methods.
\newblock {\em IEEE TVCG}, 23(1):271--280, 2017. doi: {{%
10\hspace{.1pt}\discretionary{.}{%
}{.}\hspace{.4pt}1109\discretionary{/}{%
}{/}TVCG\hspace{.1pt}\discretionary{.}{%
}{.}\hspace{.4pt}2016\hspace{.1pt}\discretionary{.}{%
}{.}\hspace{.4pt}2598544}}


\bibitem{demiralp2009coloring}
{\c{C}}.~Demiralp, J.~F. Hughes, and D.~H. Laidlaw.
\newblock Coloring {3D} line fields using {Boy’s} real projective plane
  immersion.
\newblock {\em IEEE TVCG}, 15(6):1457--1464, 2009. doi: {{%
10\hspace{.1pt}\discretionary{.}{%
}{.}\hspace{.4pt}1109\discretionary{/}{%
}{/}TVCG\hspace{.1pt}\discretionary{.}{%
}{.}\hspace{.4pt}2009\hspace{.1pt}\discretionary{.}{%
}{.}\hspace{.4pt}125}}


\bibitem{dhar2011high}
S.~Dhar, V.~Ordonez, and T.~L. Berg.
\newblock High level describable attributes for predicting aesthetics and
  interestingness.
\newblock In {\em the proceeding of CVPR}, pp. 1657--1664, 2011. doi: {{%
10\hspace{.1pt}\discretionary{.}{%
}{.}\hspace{.4pt}1109\discretionary{/}{%
}{/}CVPR\hspace{.1pt}\discretionary{.}{%
}{.}\hspace{.4pt}2011\hspace{.1pt}\discretionary{.}{%
}{.}\hspace{.4pt}5995467}}


\bibitem{gu2016mining}
Y.~Gu, C.~Wang, T.~Peterka, R.~Jacob, and S.~H. Kim.
\newblock Mining graphs for understanding time-varying volumetric data.
\newblock {\em IEEE TVCG}, 22(1):965--974, 2016. doi: {{%
10\hspace{.1pt}\discretionary{.}{%
}{.}\hspace{.4pt}1109\discretionary{/}{%
}{/}TVCG\hspace{.1pt}\discretionary{.}{%
}{.}\hspace{.4pt}2015\hspace{.1pt}\discretionary{.}{%
}{.}\hspace{.4pt}2468031}}


\bibitem{gunther2014fast}
D.~G{\"u}nther, A.~Jacobson, J.~Reininghaus, H.-P. Seidel, O.~Sorkine-Hornung,
  and T.~Weinkauf.
\newblock Fast and memory-efficient topological denoising of {2D} and {3D}
  scalar fields.
\newblock {\em IEEE TVCG}, 20(12):2585--2594, 2014. doi: {{%
10\hspace{.1pt}\discretionary{.}{%
}{.}\hspace{.4pt}1109\discretionary{/}{%
}{/}TVCG\hspace{.1pt}\discretionary{.}{%
}{.}\hspace{.4pt}2014\hspace{.1pt}\discretionary{.}{%
}{.}\hspace{.4pt}2346432}}


\bibitem{haroz2015isotype}
S.~Haroz, R.~Kosara, and S.~L. Franconeri.
\newblock Isotype visualization: Working memory, performance, and engagement
  with pictographs.
\newblock In {\em ACM CHI}, pp. 1191--1200, 2015. doi: {{%
10\hspace{.1pt}\discretionary{.}{%
}{.}\hspace{.4pt}1145\discretionary{/}{%
}{/}2702123\hspace{.1pt}\discretionary{.}{%
}{.}\hspace{.4pt}2702275}}


\bibitem{hermosilla2017physics}
P.~Hermosilla, J.~Estrada, V.~Guallar, T.~Ropinski, A.~Vinacua, and P.-P.
  V{\'a}zquez.
\newblock Physics-based visual characterization of molecular interaction
  forces.
\newblock {\em IEEE TVCG}, 23(1):731--740, 2017. doi: {{%
10\hspace{.1pt}\discretionary{.}{%
}{.}\hspace{.4pt}1109\discretionary{/}{%
}{/}TVCG\hspace{.1pt}\discretionary{.}{%
}{.}\hspace{.4pt}2016\hspace{.1pt}\discretionary{.}{%
}{.}\hspace{.4pt}2598825}}


\bibitem{hummel2010iris}
M.~Hummel, C.~Garth, B.~Hamann, H.~Hagen, and K.~I. Joy.
\newblock Iris: Illustrative rendering for integral surfaces.
\newblock {\em IEEE TVCG}, 16(6):1319--1328, 2010. doi: {{%
10\hspace{.1pt}\discretionary{.}{%
}{.}\hspace{.4pt}1109\discretionary{/}{%
}{/}TVCG\hspace{.1pt}\discretionary{.}{%
}{.}\hspace{.4pt}2010\hspace{.1pt}\discretionary{.}{%
}{.}\hspace{.4pt}173}}


\bibitem{isola2014makes}
P.~Isola, J.~Xiao, D.~Parikh, A.~Torralba, and A.~Oliva.
\newblock What makes a photograph memorable?
\newblock {\em {IEEE} TPAMI}, 36(7):1469--1482, 2014. doi: {{%
10\hspace{.1pt}\discretionary{.}{%
}{.}\hspace{.4pt}1109\discretionary{/}{%
}{/}TPAMI\hspace{.1pt}\discretionary{.}{%
}{.}\hspace{.4pt}2013\hspace{.1pt}\discretionary{.}{%
}{.}\hspace{.4pt}200}}


\bibitem{kolesar2017fractional}
I.~Koles{\'a}r, S.~Bruckner, I.~Viola, and H.~Hauser.
\newblock A fractional cartesian composition model for semi-spatial comparative
  visualization design.
\newblock {\em IEEE TVCG}, 23(1):851--860, 2017. doi: {{%
10\hspace{.1pt}\discretionary{.}{%
}{.}\hspace{.4pt}1109\discretionary{/}{%
}{/}TVCG\hspace{.1pt}\discretionary{.}{%
}{.}\hspace{.4pt}2016\hspace{.1pt}\discretionary{.}{%
}{.}\hspace{.4pt}2598870}}


\bibitem{krone2009interactive}
M.~Krone, K.~Bidmon, and T.~Ertl.
\newblock Interactive visualization of molecular surface dynamics.
\newblock {\em IEEE TVCG}, 15(6):1391--1398, 2009. doi: {{%
10\hspace{.1pt}\discretionary{.}{%
}{.}\hspace{.4pt}1109\discretionary{/}{%
}{/}TVCG\hspace{.1pt}\discretionary{.}{%
}{.}\hspace{.4pt}2009\hspace{.1pt}\discretionary{.}{%
}{.}\hspace{.4pt}157}}


\bibitem{lampe2009curve}
O.~D. Lampe, C.~Correa, K.-L. Ma, and H.~Hauser.
\newblock Curve-centric volume reformation for comparative visualization.
\newblock {\em IEEE TVCG}, 15(6):1235--1242, 2009. doi: {{%
10\hspace{.1pt}\discretionary{.}{%
}{.}\hspace{.4pt}1109\discretionary{/}{%
}{/}TVCG\hspace{.1pt}\discretionary{.}{%
}{.}\hspace{.4pt}2009\hspace{.1pt}\discretionary{.}{%
}{.}\hspace{.4pt}136}}


\bibitem{magnus2018interactive}
J.~G. Magnus and S.~Bruckner.
\newblock Interactive dynamic volume illumination with refraction and caustics.
\newblock {\em IEEE TVCG}, 24(1):984--993, 2018. doi: {{%
10\hspace{.1pt}\discretionary{.}{%
}{.}\hspace{.4pt}1109\discretionary{/}{%
}{/}TVCG\hspace{.1pt}\discretionary{.}{%
}{.}\hspace{.4pt}2017\hspace{.1pt}\discretionary{.}{%
}{.}\hspace{.4pt}2744438}}


\bibitem{marchesin2010view}
S.~Marchesin, C.-K. Chen, C.~Ho, and K.-L. Ma.
\newblock View-dependent streamlines for {3D} vector fields.
\newblock {\em IEEE TVCG}, 16(6):1578--1586, 2010. doi: {{%
10\hspace{.1pt}\discretionary{.}{%
}{.}\hspace{.4pt}1109\discretionary{/}{%
}{/}TVCG\hspace{.1pt}\discretionary{.}{%
}{.}\hspace{.4pt}2010\hspace{.1pt}\discretionary{.}{%
}{.}\hspace{.4pt}212}}


\bibitem{marino2016planar}
J.~Marino and A.~Kaufman.
\newblock Planar visualization of treelike structures.
\newblock {\em IEEE TVCG}, 22(1):906--915, 2016. doi: {{%
10\hspace{.1pt}\discretionary{.}{%
}{.}\hspace{.4pt}1109\discretionary{/}{%
}{/}TVCG\hspace{.1pt}\discretionary{.}{%
}{.}\hspace{.4pt}2015\hspace{.1pt}\discretionary{.}{%
}{.}\hspace{.4pt}2467413}}


\bibitem{marriott2012memorability}
K.~Marriott, H.~Purchase, M.~Wybrow, and C.~Goncu.
\newblock Memorability of visual features in network diagrams.
\newblock {\em IEEE TVCG}, 18(12):2477--2485, 2012. doi: {{%
10\hspace{.1pt}\discretionary{.}{%
}{.}\hspace{.4pt}1109\discretionary{/}{%
}{/}TVCG\hspace{.1pt}\discretionary{.}{%
}{.}\hspace{.4pt}2012\hspace{.1pt}\discretionary{.}{%
}{.}\hspace{.4pt}245}}


\bibitem{meyer2008glyph}
J.~Meyer-Spradow, L.~Stegger, C.~D{\"o}ring, T.~Ropinski, and K.~Hinrichs.
\newblock Glyph-based {SPECT} visualization for the diagnosis of coronary
  artery disease.
\newblock {\em IEEE TVCG}, 14(6):1499--1506, 2008. doi: {{%
10\hspace{.1pt}\discretionary{.}{%
}{.}\hspace{.4pt}1109\discretionary{/}{%
}{/}TVCG\hspace{.1pt}\discretionary{.}{%
}{.}\hspace{.4pt}2008\hspace{.1pt}\discretionary{.}{%
}{.}\hspace{.4pt}136}}


\bibitem{mirzargar2011quasi}
M.~Mirzargar and A.~Entezari.
\newblock Quasi interpolation with {Voronoi} splines.
\newblock {\em IEEE TVCG}, 17(12):1832--1841, 2011. doi: {{%
10\hspace{.1pt}\discretionary{.}{%
}{.}\hspace{.4pt}1109\discretionary{/}{%
}{/}TVCG\hspace{.1pt}\discretionary{.}{%
}{.}\hspace{.4pt}2011\hspace{.1pt}\discretionary{.}{%
}{.}\hspace{.4pt}230}}


\bibitem{mohammed2018abstractocyte}
H.~Mohammed, A.~K. Al-Awami, J.~Beyer, C.~Cali, P.~Magistretti, H.~Pfister, and
  M.~Hadwiger.
\newblock Abstractocyte: A visual tool for exploring nanoscale astroglial
  cells.
\newblock {\em IEEE TVCG}, 24(1):853--861, 2018. doi: {{%
10\hspace{.1pt}\discretionary{.}{%
}{.}\hspace{.4pt}1109\discretionary{/}{%
}{/}TVCG\hspace{.1pt}\discretionary{.}{%
}{.}\hspace{.4pt}2017\hspace{.1pt}\discretionary{.}{%
}{.}\hspace{.4pt}2744278}}


\bibitem{quan2018intelligent}
T.~M. Quan, J.~Choi, H.~Jeong, and W.-K. Jeong.
\newblock An intelligent system approach for probabilistic volume rendering
  using hierarchical {3D} convolutional sparse coding.
\newblock {\em IEEE TVCG}, 24(1):964--973, 2018. doi: {{%
10\hspace{.1pt}\discretionary{.}{%
}{.}\hspace{.4pt}1109\discretionary{/}{%
}{/}TVCG\hspace{.1pt}\discretionary{.}{%
}{.}\hspace{.4pt}2017\hspace{.1pt}\discretionary{.}{%
}{.}\hspace{.4pt}2744078}}


\bibitem{rautek2014vislang}
P.~Rautek, S.~Bruckner, M.~E. Gr{\"o}ller, and M.~Hadwiger.
\newblock {ViSlang}: A system for interpreted domain-specific languages for
  scientific visualization.
\newblock {\em IEEE TVCG}, 20(12):2388--2396, 2014. doi: {{%
10\hspace{.1pt}\discretionary{.}{%
}{.}\hspace{.4pt}1109\discretionary{/}{%
}{/}TVCG\hspace{.1pt}\discretionary{.}{%
}{.}\hspace{.4pt}2014\hspace{.1pt}\discretionary{.}{%
}{.}\hspace{.4pt}2346318}}


\bibitem{rosenholtz2005feature}
R.~Rosenholtz, Y.~Li, J.~Mansfield, and Z.~Jin.
\newblock Feature congestion: a measure of display clutter.
\newblock In {\em ACM CHI}, pp. 761--770. ACM, 2005. doi: {{%
10\hspace{.1pt}\discretionary{.}{%
}{.}\hspace{.4pt}1145\discretionary{/}{%
}{/}1054972\hspace{.1pt}\discretionary{.}{%
}{.}\hspace{.4pt}1055078}}


\bibitem{rosenholtz2007measuring}
R.~Rosenholtz, Y.~Li, and L.~Nakano.
\newblock Measuring visual clutter.
\newblock {\em Journal of Vision}, 7(2):17, 2007. doi: {{%
10\hspace{.1pt}\discretionary{.}{%
}{.}\hspace{.4pt}1167\discretionary{/}{%
}{/}7\hspace{.1pt}\discretionary{.}{%
}{.}\hspace{.4pt}2\hspace{.1pt}\discretionary{.}{%
}{.}\hspace{.4pt}17}}


\bibitem{sagrista2017topological}
A.~Sagrista, S.~Jordan, A.~Just, F.~Dias, L.~G. Nonato, and F.~Sadlo.
\newblock Topological analysis of inertial dynamics.
\newblock {\em IEEE TVCG}, 23(1):950--959, 2017. doi: {{%
10\hspace{.1pt}\discretionary{.}{%
}{.}\hspace{.4pt}1109\discretionary{/}{%
}{/}TVCG\hspace{.1pt}\discretionary{.}{%
}{.}\hspace{.4pt}2016\hspace{.1pt}\discretionary{.}{%
}{.}\hspace{.4pt}2599018}}


\bibitem{saket2016comparing}
B.~Saket, C.~Scheidegger, and S.~Kobourov.
\newblock Comparing node-link and node-link-group visualizations from an
  enjoyment perspective.
\newblock In {\em CGF}, vol.~35, pp. 41--50, 2016. doi: {{%
10\hspace{.1pt}\discretionary{.}{%
}{.}\hspace{.4pt}1111\discretionary{/}{%
}{/}cgf\hspace{.1pt}\discretionary{.}{%
}{.}\hspace{.4pt}12880}}


\bibitem{schlegel2011extinction}
P.~Schlegel, M.~Makhinya, and R.~Pajarola.
\newblock Extinction-based shading and illumination in {GPU} volume
  ray-casting.
\newblock {\em IEEE TVCG}, 17(12):1795--1802, 2011. doi: {{%
10\hspace{.1pt}\discretionary{.}{%
}{.}\hspace{.4pt}1109\discretionary{/}{%
}{/}TVCG\hspace{.1pt}\discretionary{.}{%
}{.}\hspace{.4pt}2011\hspace{.1pt}\discretionary{.}{%
}{.}\hspace{.4pt}198}}


\bibitem{tamura1978textural}
H.~Tamura, S.~Mori, and T.~Yamawaki.
\newblock Textural features corresponding to visual perception.
\newblock {\em IEEE TSMC}, 8(6):460--473, 1978. doi: {{%
10\hspace{.1pt}\discretionary{.}{%
}{.}\hspace{.4pt}1109\discretionary{/}{%
}{/}TSMC\hspace{.1pt}\discretionary{.}{%
}{.}\hspace{.4pt}1978\hspace{.1pt}\discretionary{.}{%
}{.}\hspace{.4pt}4309999}}


\bibitem{tao2012structure}
Y.~Tao, H.~Lin, F.~Dong, C.~Wang, G.~Clapworthy, and H.~Bao.
\newblock Structure-aware lighting design for volume visualization.
\newblock {\em IEEE TVCG}, 18(12):2372--2381, 2012. doi: {{%
10\hspace{.1pt}\discretionary{.}{%
}{.}\hspace{.4pt}1109\discretionary{/}{%
}{/}TVCG\hspace{.1pt}\discretionary{.}{%
}{.}\hspace{.4pt}2012\hspace{.1pt}\discretionary{.}{%
}{.}\hspace{.4pt}267}}


\bibitem{thomas2013detecting}
D.~M. Thomas and V.~Natarajan.
\newblock Detecting symmetry in scalar fields using augmented extremum graphs.
\newblock {\em IEEE TVCG}, 19(12):2663--2672, 2013. doi: {{%
10\hspace{.1pt}\discretionary{.}{%
}{.}\hspace{.4pt}1109\discretionary{/}{%
}{/}TVCG\hspace{.1pt}\discretionary{.}{%
}{.}\hspace{.4pt}2013\hspace{.1pt}\discretionary{.}{%
}{.}\hspace{.4pt}148}}


\bibitem{thomas2014multiscale}
D.~M. Thomas and V.~Natarajan.
\newblock Multiscale symmetry detection in scalar fields by clustering
  contours.
\newblock {\em IEEE TVCG}, 20(12):2427--2436, 2014. doi: {{%
10\hspace{.1pt}\discretionary{.}{%
}{.}\hspace{.4pt}1109\discretionary{/}{%
}{/}TVCG\hspace{.1pt}\discretionary{.}{%
}{.}\hspace{.4pt}2014\hspace{.1pt}\discretionary{.}{%
}{.}\hspace{.4pt}2346332}}


\bibitem{viola2017pondering}
I.~Viola and T.~Isenberg.
\newblock Pondering the concept of abstraction in (illustrative) visualization.
\newblock {\em IEEE TVCG (pre-print)}, 2018. doi: {{%
10\hspace{.1pt}\discretionary{.}{%
}{.}\hspace{.4pt}1109\discretionary{/}{%
}{/}TVCG\hspace{.1pt}\discretionary{.}{%
}{.}\hspace{.4pt}2017\hspace{.1pt}\discretionary{.}{%
}{.}\hspace{.4pt}2747545}}


\bibitem{von2008smoke}
W.~Von~Funck, T.~Weinkauf, H.~Theisel, and H.-P. Seidel.
\newblock Smoke surfaces: An interactive flow visualization technique inspired
  by real-world flow experiments.
\newblock {\em IEEE TVCG}, 14(6):1396--1403, 2008. doi: {{%
10\hspace{.1pt}\discretionary{.}{%
}{.}\hspace{.4pt}1109\discretionary{/}{%
}{/}TVCG\hspace{.1pt}\discretionary{.}{%
}{.}\hspace{.4pt}2008\hspace{.1pt}\discretionary{.}{%
}{.}\hspace{.4pt}163}}


\bibitem{wang2011information}
C.~Wang and H.-W. Shen.
\newblock Information theory in scientific visualization.
\newblock {\em Entropy}, 13(1):254--273, 2011. doi: {{%
10\hspace{.1pt}\discretionary{.}{%
}{.}\hspace{.4pt}3390\discretionary{/}{%
}{/}e13010254}}


\bibitem{wu2017direct}
K.~Wu, A.~Knoll, B.~J. Isaac, H.~Carr, and V.~Pascucci.
\newblock Direct multifield volume ray casting of fiber surfaces.
\newblock {\em IEEE TVCG}, 23(1):941--949, 2017. doi: {{%
10\hspace{.1pt}\discretionary{.}{%
}{.}\hspace{.4pt}1109\discretionary{/}{%
}{/}TVCG\hspace{.1pt}\discretionary{.}{%
}{.}\hspace{.4pt}2016\hspace{.1pt}\discretionary{.}{%
}{.}\hspace{.4pt}2599040}}


\bibitem{zhang2013lighting}
Y.~Zhang and K.-L. Ma.
\newblock Lighting design for globally illuminated volume rendering.
\newblock {\em IEEE TVCG}, 19(12):2946--2955, 2013. doi: {{%
10\hspace{.1pt}\discretionary{.}{%
}{.}\hspace{.4pt}1109\discretionary{/}{%
}{/}TVCG\hspace{.1pt}\discretionary{.}{%
}{.}\hspace{.4pt}2013\hspace{.1pt}\discretionary{.}{%
}{.}\hspace{.4pt}172}}


\bibitem{zhao2017bivariate}
H.~Zhao and J.~Chen.
\newblock Bivariate separable-dimension glyphs can improve visual analysis of
  holistic features.
\newblock {\em arXiv preprint arXiv:1712.02333}, 2017.

\end{thebibliography}
\end{document}